\documentclass[aps,pra,floatfix,showpacs,twocolumn,tightenlines,superscriptaddress]{revtex4}

\usepackage{amsmath}
\usepackage{graphicx}

\begin{document}

\title {Quantum computation with optical coherent states}
\author
{T.C.Ralph, A.Gilchrist, G.J.Milburn}
\affiliation {Centre for Quantum Computer Technology,
Department of Physics, University of Queensland Brisbane, QLD 4072,
Australia\\
email: ralph@physics.uq.edu.au}
\author {W.J.Munro}
\affiliation {Hewlett Packard Laboratories,
Filton Road Stoke Gifford,
Bristol BS34 8QZ, U.K}

\author {S.Glancy}
\affiliation {Department of Physics, University of Notre Dame,
Notre Dame, Indiana 46556, USA}

\begin{abstract}
We show that quantum computation circuits using coherent states as the
logical qubits can be constructed from simple
linear networks, conditional photon measurements and ``small''
coherent superposition resource states.

\end{abstract}

\maketitle

\vspace*{10pt}
%

\section{Introduction}

Quantum optics has proved a fertile field for experimental tests
of quantum information science.
However, quantum optics was not thought to provide a practical
path to efficient and scalable
quantum computation.
This orthodoxy was challenged when
Knill et al.\cite{KLM} showed that, given single photon sources
and single photon detectors, linear optics alone would suffice
to implement efficient quantum computation. While this result is
surprising, the complexity of the optical networks required is
daunting.

More recently it has become clear that other, quite different versions
of this paradigm are possible. In particular, by encoding the quantum
information in multi-photon
coherent states, rather than single photon states,
an efficient scheme which is elegant in its
simplicity has been proposed \cite{RMM}. The required resource, which
may be produced non-deterministically, is a superposition of coherent
states. Given this, the scheme is deterministic and
requires only relatively simple linear optical networks and photon
counting.
Unfortunately the amplitude of the required resource states is
prohibitively large. Here we build on this idea and show that with
only a moderate increase in complexity a
scheme based on much smaller superposition states is possible.

The idea of encoding quantum information on continuous variables
of multi-photon fields \cite{Bra98}
has led to a number of proposals for realizing quantum computation in this way
\cite{llo,sand,kim01}. One drawback of these proposals is that
``hard'', non-linear interactions are required ``in-line'' of the
computation. These would be very
difficult to implement in practice. In contrast, this proposal
requires only ``easy', linear in-line interactions. The hard
interactions are only required for ``off-line'' production of resource
states. A related proposal is that of Gottesman et al \cite{pre} in
which superpositions of squeezed states are used to encode the qubits.
In that proposal the hard interactions are only used for the initial state
preparation.
However, quadratic, squeezing type interactions, are required in-line
along with linear interactions.

This paper is laid out in the following way. We start by describing
the basic principles of the scheme. In sections
III and IV we describe realistic measurement and resource production
techniques respectively, based on photon counting and linear optics.
In section V we consider error correction and we conclude in Section
VI.

\section{Basic Scheme}

The output of a single mode, stabilized laser can be described by a
coherent state, $|\alpha \rangle$, where $\alpha$ is a complex number
which
determines the average field amplitude.
Coherent states are defined by unitary transformation of the vacuum
\cite{WallsMilb94}, $|\alpha\rangle=D(\alpha)|0\rangle$, where
$D(\alpha)$ is the displacement operator.
Let us consider an encoding of logical qubits in coherent states
with $|0\rangle_L \equiv
    |-\alpha \rangle$ and
$|1\rangle_L  \equiv  |\alpha\rangle$,
where we take $\alpha$ to be real \cite{note1}.
These qubits are not exactly orthogonal, but the approximation of
orthogonality is good for $\alpha$ even moderately large as
$|\langle \alpha | -\alpha \rangle|^{2} =e^{-4 \alpha^2}$.
We will assume for most of this paper that $\alpha \ge 2$, which
gives $|\langle \alpha | -\alpha \rangle|^{2} \le 1.1 \times 10^{-7}$.
Measurement of the qubit values can be achieved with high efficiency
by homodyne detection with respect to a local oscillator phase reference.

Of course, if one wished, an exactly orthogonal qubit code
can easily be defined in terms of the orthogonal parity eigenstates,
$\widetilde{|0\rangle}_L=|\alpha\rangle+|-\alpha\rangle\ \ , \
\widetilde{ |1\rangle}_L=|\alpha\rangle-|-\alpha\rangle$. However
such states are only a single (nonunitary)  qubit gate away from the
code we propose to use. The issue is not so much the
orthogonality
of the qubit code, but rather the need to work outside the qubit
space during qubit processing. As we shall now show, this can be done
with negligible error.

{\bf Bit-flip Gate}. The
logical value of a qubit can be flipped by delaying it with respect to
the local oscillator by half a cycle. Thus the $X$ or
``bit-flip''
gate is given by
\begin{equation}
X = \exp\{i \pi \hat a^{\dagger} \hat a \}
\end{equation}
This is a unitary gate. As already noted, the Hadamard gate (or its
equivalents)
which effects transformation from $|x\rangle_L$ to
$\widetilde{|x\rangle}_L$ cannot be unitary. This is because the
logical basis states are not orthogonal but the states
$\widetilde{|x\rangle}_L$ are parity eigenstates which are
orthogonal. For this reason we  now consider nonunitary gates based
on projective measurement. Gates based on projective
measurements  will be probabilistic in their operation.

{\bf Sign-flip Gate}. A
bit flip in the superposition basis, ie a ``sign flip'' or $Z$ gate,
can be achieved via teleportation \cite{bennett} as follows. A
resource
of coherent superposition states (commonly referred to as ``cat''
states),
$1/\sqrt{2}(|-\sqrt{2}\;\alpha \rangle +
|\sqrt{2}\;\alpha \rangle)$, is required.
Splitting such a cat state on a 50:50 beamsplitter produces the
entangled Bell state, $1/\sqrt{2}(|-\alpha, -\alpha \rangle + |\alpha, \alpha
\rangle)$. A Bell basis measurement is then made on the qubit state,
$\mu|-\alpha \rangle + \nu |\alpha \rangle$, and one half of the
entangled state (where $\mu$ and $\nu$ are arbitrary complex numbers).
Depending on which of the four possible outcomes are
found the other half of the Bell state is projected into one of the
following four
states with equal probability:
\begin{eqnarray}
& &     \mu|-\alpha \rangle + \nu |\alpha \rangle \nonumber\\
& &     \mu|-\alpha \rangle - \nu |\alpha \rangle \nonumber\\
& &     \mu|\alpha \rangle + \nu |-\alpha \rangle \nonumber\\
& &     \mu|\alpha \rangle - \nu |-\alpha \rangle
\end{eqnarray}
The bit flip errors in results three and four can be corrected using the
$X$ gate. After $X$ correction the gate has two possible outcomes:
either the identity has been applied, in which case we repeat the
process, or else the required transformation
\begin{equation}
Z (\mu|-\alpha \rangle + \nu |\alpha \rangle) =
\mu|-\alpha \rangle - \nu |\alpha \rangle
\end{equation}
has been implemented. On average this will take
two attempts. We write
\begin{equation}
Z = T_{X}^{p}
\end{equation}
meaning the teleportation transformation, $T$, with bit-flip
correction, $X$, is applied $p$ times, where $p$ is outcome dependent.

Our remaining gates implement operations which
may conveniently be described by the product
operator notation
\begin{align}
R(K_{i} \otimes
    K_{j},\theta) &=e^{-i{{\theta}\over{2}}K_{i} \otimes
    K_{j}}\\ &=
\cos(\theta/2)I \otimes I - i\sin(\theta/2) K_{i} \otimes
    K_{j}
\label{nom}
\end{align}
where $K_{i,j}$ can take on the values, $X$,
$Y$, $Z$ or $I$ (the Pauli sigma operators and the identity).
For single qubit operations we will drop the
redundant identity ($I$) operation on the second qubit.

{\bf Phase Rotation Gate}. Consider an arbitrary single qubit
rotation about $Z$,
$R(Z,\theta)$. This can be implemented by shifting our qubit a small
distance out of the computational basis and then using teleportation
to project back. We begin by displacing our arbitrary input qubit by a
small amount in the imaginary direction (see
Fig.1(a))
\begin{figure}
   \begin{tabular}{cc}
     (a) & \includegraphics[scale=.8]{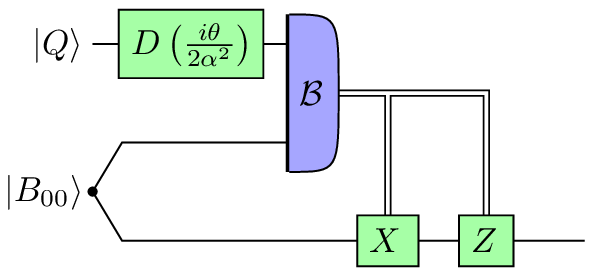} \\
     (b) & \includegraphics[scale=.7]{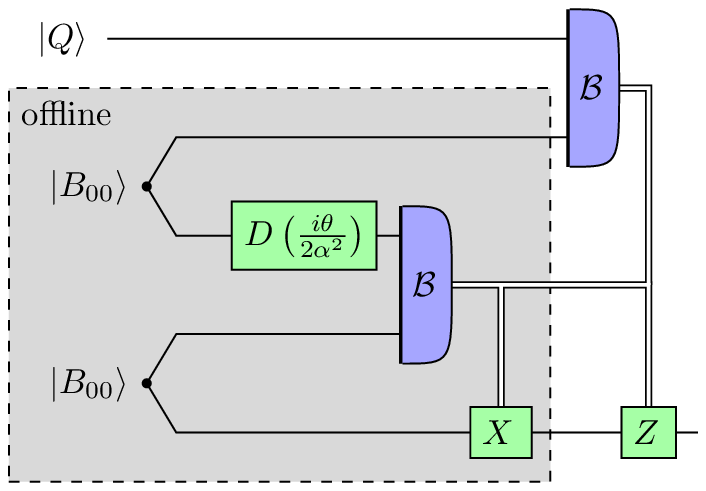} 
   \end{tabular}
\caption{Schematics of implementing the $R(Z,\theta)$ gate.
   (a) The bare gate; its operation is near deterministic for a
   sufficiently small value of $\theta/\alpha$. Repeated application of
   this gate can build up a finite rotation with high probability.  (b)
   The teleported gate; its operation is deterministic, however it may
   need to be applied several times in order to achieve the correct
   rotation. Determinism is achieved by preparing the entangled
   resource ``offline'' and only applying the gate to the qubit when
   the resource is available. In the diagrams, $\mathcal{B}$ represents
   a cat-Bell measurement.}
\label{fig:fid210}
\end{figure}
\begin{multline}
D\left(\frac{i\theta}{2\alpha^{2}}\right)(\mu|-\alpha \rangle + \nu
|\alpha \rangle)  = \\
\mu|-\alpha\left(1-\frac{i\theta}{2\alpha^2}\right) \rangle + \nu
|\alpha\left(1+\frac{i\theta}{2\alpha^2}\right) \rangle
\end{multline}
Now consider the effect of teleporting this state.  Using
the relationship \cite{WallsMilb94}
\begin{equation}
         \langle \tau|\alpha \rangle =
\exp[-1/2(|\tau|^{2}+|\alpha|^{2})+\tau^{*} \alpha]
\label{overlap}
\end{equation}
we find the
required projections are approximately given by
\begin{eqnarray}
\langle \pm \alpha | \pm \alpha\left(1 \pm
\frac{i\theta}{2\alpha^2}\right) \rangle & = &
e^{\pm i \theta/2} e^{-\frac{\theta^2}{8\alpha^2}} \nonumber\\
\langle \mp \alpha | \pm \alpha\left(1 \pm
\frac{i\theta}{2\alpha^2}\right) \rangle & = & 0
\end{eqnarray}
where we have assumed orthogonality and that
$\frac{\theta^2}{8\alpha^2}<<1$. The outcome of the sequence of
displacement followed
by teleportation is then found to be
\begin{multline}
T_{X} D\left(\frac{i\theta}{2\alpha^2}\right)(\mu|-\alpha \rangle +
\nu |\alpha \rangle) =\\
e^{-\frac{\theta^2}{8\alpha^2}}
(e^{-i \theta/2}\mu|-\alpha \rangle \pm e^{i \theta/2}
\nu |\alpha \rangle)
\label{RZ}
\end{multline}
The ``$\pm$'' sign depends on the Bell state measurement outcome and can
be corrected by the $Z$ gate. The transformation
is then $R(Z,\theta)$.

Notice however, that
the output state in Eq.\ref{RZ} is unnormalized. This reflects the
fact that, because we are projecting back onto the qubit basis from
outside, the probability of success is not unity. In
other words there is a probability, $P=1-e^{{{-\theta^{2}}\over{4
\alpha^{2}}}}$,
that the Bell state measurement will return a null result, in which
case the gate will fail. In order to make the probability of failure
as small as possible we require $\theta^{2} << 4 \alpha^{2}$.
One option would be to
let $\alpha$ be large \cite{RMM}.
In this way $\theta$ can be a significant angle
whilst $P \approx 1$ is still satisfied. However, this is
undesirable because of the difficulty in producing cat states with
large $\alpha$.

A second option is to implement the gate with an
incremental phase shift, repeatedly, to build up a significant angle.
Let $\theta=n\phi$, then after $n$ rotations by $\phi$ we have
\begin{multline}
(T_{X} D(\frac{i\phi}{2\alpha^{2}}))^{n}(\mu|-\alpha \rangle + \nu
|\alpha \rangle) =\\
e^{-\frac{n \phi^{2}}{8\alpha^{2}}}
(e^{-i n \phi/2}\mu|-\alpha \rangle \pm e^{i n\phi/2}
\nu |\alpha \rangle)
\label{RZi}
\end{multline}
The transformation is again $R(Z,\theta)$. The success probability is
$P=e^{{{-\theta^{2}}\over{4 n \alpha^{2}}}}$, which can be made
arbitrarily close to one for small $\alpha$ simply by choosing $n$
sufficiently large. For example with $\alpha = 2$, $\theta=\pi/4$ and
$n=8$ we find $P=0.995$. (or $n=30$ gives $P=.999$).
This is basically an application of the quantum Zeno effect
\cite{zeno}.

A third option is to use the technique of gate
teleportation \cite{gott}. In this case we place the gate inside a
second teleporter as shown schematically in Fig.1(b). The
$R(Z,\theta)$
gate of Eq.\ref{RZ} is implemented on one arm of a second Bell-cat
state. If (and only if) the gate is successful, a Bell measurement
is made between the qubit and the other arm of the entangled state. It is
straightforward to show that the output state
after $X$ and $Z$ correction is
\begin{equation}
e^{\mp i \theta/2}\mu|-\alpha \rangle + e^{\pm i \theta/2}
\nu |\alpha \rangle
\label{RZT}
\end{equation}
The signs in the arguments of the exponentials depend on the Bell state
measurement results. The qubit is teleported
with an equal probability of either $R(Z,\theta)$ or $R(Z,-\theta)$
applied. The operation is deterministic for the qubit as the
second teleportation is only carried through if the first one is
successful. In general the result $R(Z,-\theta)$ can be corrected by
applying the gate again, but this time attempting to apply $R(Z,2
\theta)$.
If this again fails the the process is continued by attempting to
apply $R(Z,4 \theta)$ etc. Symmetry can be exploited for
certain angles. For example for the ``phase'' gate, $R(Z, \pi/2)$,
only $X$ and $Z$ corrections are necessary.

{\bf Controlled Phase Gate}. A non-trivial 2-qubit gate,
$R(Z\otimes Z,-\phi)$, can be
implemented in a similar way to the single qubit rotation (see
Fig.2(a)).
Consider the beamsplitter
interaction given by the unitary transformation
\begin{equation}
U_{ab}=\exp[i {{\theta}\over{2}} (a b^{\dagger}+a^{\dagger} b)]
\end{equation}
where $a$ and $b$ are the annihilation operators corresponding to two
coherent state qubits $|\gamma \rangle_{a}$ and $|\beta
\rangle_{b}$, with $\gamma$ and $\beta$ taking values of $-\alpha$ or
$\alpha$. It is well known that the output state
produced by such an interaction is
\begin{eqnarray}
U_{ab} |\gamma \rangle_{a} |\beta \rangle_{b}=|\cos
{{\theta}\over{2}} \gamma+i
\sin {{\theta}\over{2}} \beta \rangle_{a} |\cos {{\theta}\over{2}}
\beta+
i \sin {{\theta}\over{2}} \gamma \rangle_{b}
    \label{Ho}
\end{eqnarray}
where $\cos^{2} {{\theta}\over{2}}$ ($\sin^{2} {{\theta}\over{2}}$)
is the reflectivity
(transmissivity) of the beamsplitter. If both output beams are now
projected using teleportation as for the single qubit gate we find
for an arbitrary input state
\begin{multline}
 T_{Xa} T_{Xb} U_{ab}(\nu|-\alpha \rangle_{a} |-\alpha \rangle_{b}+
\mu|\alpha \rangle_{a} |-\alpha \rangle_{b}+\\ 
\tau|-\alpha
\rangle_{a} |\alpha \rangle_{b}+\gamma|\alpha \rangle_{a} |\alpha
\rangle_{b}) \\
 \; \; \; \; = e^{-\theta^{2} \alpha^{2}/4}
    (e^{i \theta \alpha^{2}}\nu|-\alpha \rangle_{a} |-\alpha
    \rangle_{b}\pm
e^{-i \theta \alpha^{2}}\mu|\alpha \rangle_{a} |-\alpha \rangle_{b}
\pm \\
\; \; \; \; \; \; \; \; \; \; \; e^{-i \theta
\alpha^{2}}\tau|-\alpha
\rangle_{a} |\alpha \rangle_{b}+
e^{i \theta \alpha^{2}}
\gamma|\alpha \rangle_{a} |\alpha \rangle_{b}
    \label{Ho1}
\end{multline}
where as before we have assumed orthogonality and that $\theta^{2}
\alpha^{2}<<1$ and the $\pm$ signs depend on the outcome of the
Bell
measurements. If we choose $\phi=2 \theta \alpha^{2}=\pi/2$ then
$R(Z\otimes Z,-\pi/2)$ is implemented, a gate that can easily be
shown to be locally equivalent to a CNOT.

Once again the probability of success is non-unit and two options are
possible for small $\alpha$: repeated iterations of the gate for an
incremental value of $\phi$ can be used to build up to a total angle
of $\pi/2$ with a high probability of success via the quantum zeno
effect or; we
can use gate teleportation to guarantee success. To achieve the
second gate teleportation we must now nest the two qubit
gate inside two teleporters as shown schematically in Fig.2(b). Only
$X$ and $Z$ corrections are required.
\begin{figure}
   \begin{tabular}{cc}
     (a) & \includegraphics[scale=.8]{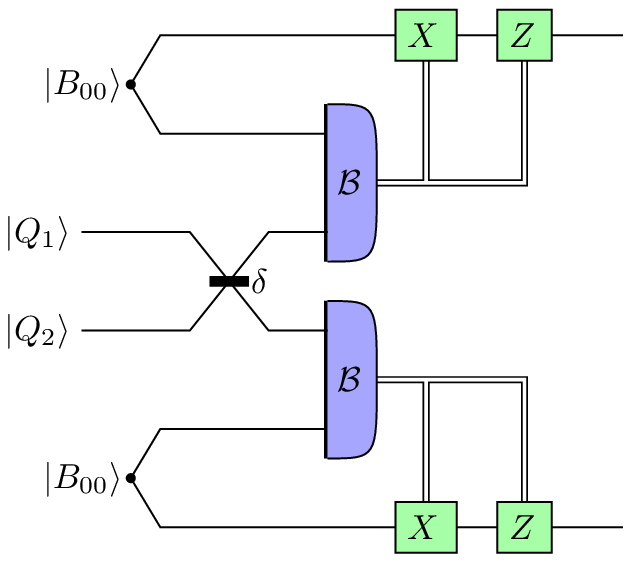}\\
     (b) & \includegraphics[scale=.7]{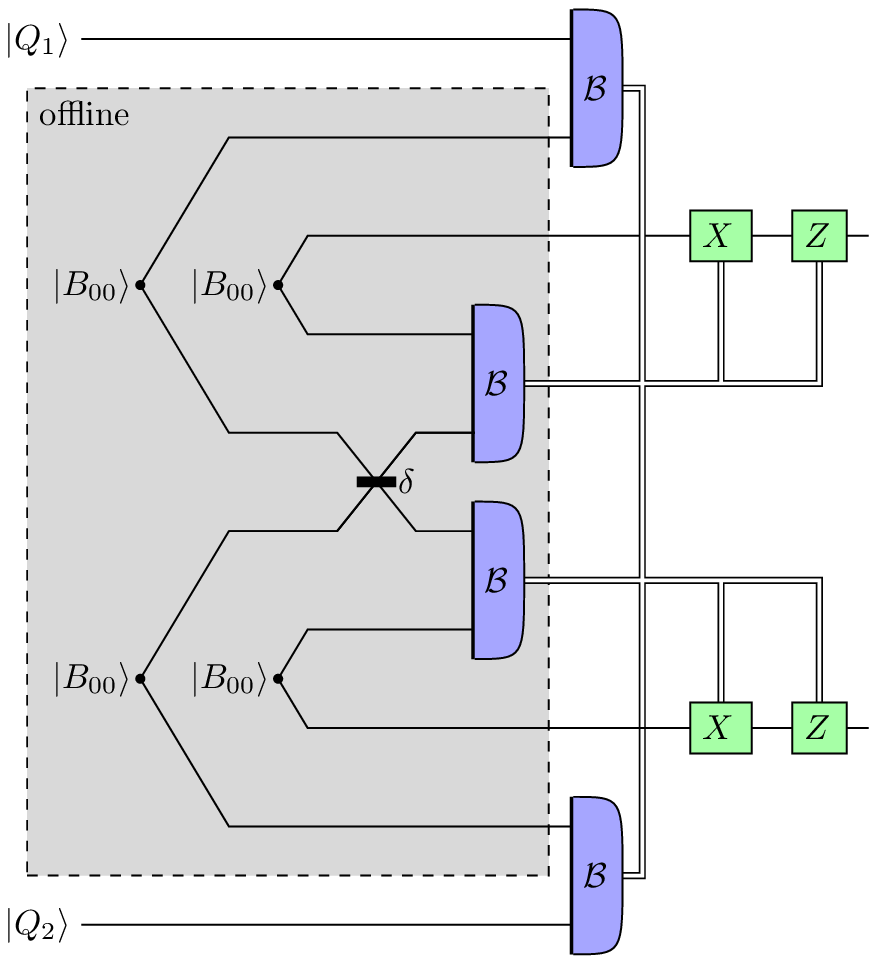}
   \end{tabular}
\caption{Schematics of implementing the $R(Z \otimes Z,-\pi/2)$ gate.
   (a) The bare gate; its operation is near deterministic for a
   sufficiently small value of $\theta^{2} \alpha^{2}$ where the
   refectivity of the beamsplitter is $\delta=\cos^{2} {{\theta}\over{2}}$.
   Repeated
   application of this gate can build up to a $\pi/2$ rotation with
   high probability.  (b) The teleported gate; its operation is
   deterministic.  Determinism is achieved by preparing the entangled
   resource ``offline'' and only applying the gate to the qubits when
   the resource is available. In the diagrams, $\mathcal{B}$ represents
   a cat-Bell measurement.}
\label{fig:fig2}
\end{figure}

{\bf Superposition Gate}. To complete our set of gates we now describe
how to implement a rotation of $\pi/2$
about $X$, ie $R(X,\pi/2)$. This gate takes computational basis qubits
into the diagonal, or superposition, basis and is locally equivalent
to a Hadamard gate. The gate is shown schematically in
Fig.3(a).
It is similar to the $Z$ rotation except now the displacement followed
by Bell state measurement on the qubit and one of the Bell state modes is
replaced
by the beamsplitter interaction used in the $R(Z\otimes Z,-\pi/2)$
gate,
followed by single (as opposed to Bell-) cat measurements on the
output modes from the beamsplitter. The interaction produces the
following output state from an arbitrary input
\begin{multline}
C_{a}C_{b}U_{BS}(\mu |-\alpha \rangle + \nu |\alpha \rangle) =\\
e^{-\theta^{2}\alpha^{2}/4}\{(e^{i \theta \alpha^{2}}\mu \pm
e^{-i \theta \alpha^{2}}\nu) |-\alpha \rangle \\+
(\pm e^{-i \theta \alpha^{2}}\mu \pm
e^{i \theta \alpha^{2}}\nu) |\alpha \rangle \}
\end{multline}
where $C_{a}$ and $C_{b}$ represent the cat state projections. The
$\pm$ signs depend on the outcome of the cat state measurements.
Using $X$ and $Z$ gates we can correct all the $\pm$'s to $+$'s.
Choosing $2 \theta \alpha^{2}=\pi/2$ then implements $R(X,\pi/2)$. As
before the gate is probabilistic for small $\alpha$,
working with a probability of
$e^{-\theta^{2}\alpha^{2}/2}$. To achieve near determinism using the
quantum zeno effect one would replace the beamsplitter interaction
(within the dashed box of Fig.3(a))
with the $R(Z\otimes Z,-\phi)$ gate of Fig.2(a), iterated sufficient
times to give $\phi= \pi/2$ with high probability of success. The
rest of gate remains the same and will work deterministically.
As before we can also implement the gate
deterministically using gate teleportation as depicted in Fig.3(b).
Only
$X$ and $Z$ corrections are required.
\begin{figure}
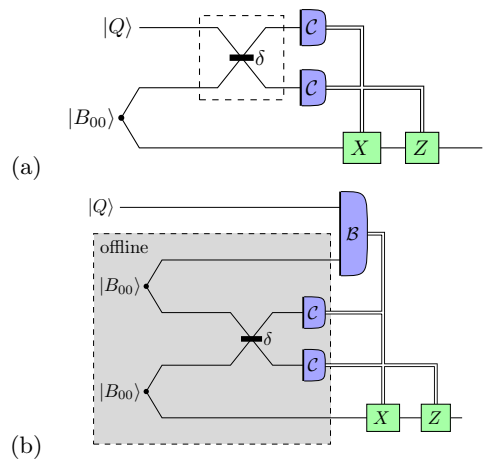

   \begin{tabular}{cc}
     (a) & \includegraphics[scale=.8]{RX_circ.eps}\\
     (b) &\includegraphics[scale=.7]{RXtel_circ.eps}
   \end{tabular}
\caption{Schematics of implementing the $R(X,\pi/2)$ gate.
   (a) The bare gate; its operation is near deterministic for a
   sufficiently small value of $\theta^{2} \alpha^{2}$. Replacement of
   the dashed section with the repeated application of the gate of
   Fig.2(a) can build up to a $R(X,\pi/2)$ rotation with high
   probability.  (b) The teleported gate; its operation is
   deterministic.  Determinism is achieved by preparing the entangled
   resource ``offline'' and only applying the gate to the qubits when
   the resource is available. In the diagrams, $\mathcal{B}$ represents
   a cat-Bell measurement, and $\mathcal{C}$ represents a cat
   measurement.}
\label{fig:fig3}
\end{figure}

The gates $R(Z,\theta)$, $R(X,\pi/2)$ and $R(Z\otimes Z,-\pi/2)$ form a
universal set. An arbitrary single qubit rotation can be constructed
from $R(Z,\psi) R(X,\pi/2) R(Z,\phi) R(X,-\pi/2)$ and as commented
before $R(Z\otimes Z,-\pi/2)$ is locally equivalent to a CNOT. This
completes our basic discussion. In the next section we consider how
the required cat and Bell state measurements can be performed.

\section{Cat-Basis Measurements}

We define a cat basis measurement to be some procedure that projects
the state of an optical mode onto one of the two states
$\frac{1}{\sqrt{2}}\left(|-\alpha\rangle\pm|\alpha\rangle\right)$.
If our input state consists only of an arbitrary superposition of
these 2 states then cat-basis measurement can be achieved by simply
counting the photons in the mode.
An even number of detected photons indicates measurement of
the
state
$\frac{1}{\sqrt{2}}\left(|-\alpha\rangle+|\alpha\rangle\right)$,
and an odd number of photons indicates measurement of
$\frac{1}{\sqrt{2}}\left(|-\alpha\rangle-|\alpha\rangle\right)$.  Of
course, this will require very high quality photon detectors which
can
reliably distinguish $n$ from $n+1$ photons when $n\sim\alpha^2$.

The cat states can also be distinguished to some extent by homodyne
detection
looking at the imaginary quadrature.  Cat states display fringes in
the imaginary quadrature which are $\pi/2$ out of phase between the
plus and minus cats \cite{reid}.  Therefore a measurement result that
falls close to a fringe maximum can be identified with one or other
cat with high probability.  This technique gives inconclusive results
some of the time (i.e. close to the fringe crossings) but could prove
useful for initial experimental demonstrations.

In order to perform a Bell basis measurement on two modes (say, modes
$a$ and $b$) containing coherent state qubits we can employ the
following
procedure  \cite{enk02} \cite{kim2}.
Allow the two qubits to interfere at a 50:50 beam splitter
$B_{a,b}=\exp\left[\frac{\pi}{4}\left(-a^{\dagger} b+a
b^{\dagger}\right)\right]$, where $a$ and $b$ are the
annihilation operators for modes $a$ and $b$.  Then use photon
counters to measure the number of photons in each mode.  We can then
identify the four possible results:
\begin{enumerate}
\item an even number of photons in mode $a$ and zero photons in mode
$b$,
\item an odd number of photons in mode $a$ and zero photons in mode
$b$,
\item zero photons in mode $a$ and an even number of photons in mode
$b$,
or \item zero photons in mode $a$ and an odd number of photons in mode
$b$;
\end{enumerate}
corresponding to each of the four Bell-cat states:
\begin{enumerate}
\item
$|B_{00}\rangle=\frac{1}{\sqrt{2}}\left(|-\alpha,-\alpha\rangle+
|\alpha,\alpha\rangle\right)$,

\item
$|B_{10}\rangle=\frac{1}{\sqrt{2}}\left(|-\alpha,-\alpha\rangle-
|\alpha,\alpha\rangle\right)$,

\item
$|B_{01}\rangle=\frac{1}{\sqrt{2}}\left(|-\alpha,\alpha\rangle+
|\alpha,-\alpha\rangle\right)$,

or \item
$|B_{11}\rangle=\frac{1}{\sqrt{2}}\left(|-\alpha,\alpha\rangle-
|\alpha,-\alpha\rangle\right)$.

\end{enumerate}
Note that there is also a fifth possibility of detecting zero photons
in both modes $a$ and $b$, which indicates a failure of the
measurement.
Fortunately, this occurs with probability of only $\sim
e^{-\alpha^2}$.  The preceding discussion assumed we were only
differentiating between states within the computational basis.
However, the gates discussed in section II involved moving short
distances outside this basis.  Nevertheless we will show in the
following that these types of measurements are sufficient to implement
our gates.

As an example, we will examine the use of this procedure for the Bell
state measurement required when performing $R(Z,\theta)$.  In order to
perform this rotation, we must use the displacement
$D(\frac{i\theta}{2\alpha^{2}})$ on the qubit $|Q\rangle$ in mode $a$ and
append the
Bell state
$\frac{1}{\sqrt{2}}\left(|-\alpha,-\alpha\rangle+|\alpha,\alpha\rangle\right)$
in modes $b$ and $c$.  When modes $a$ and $b$ meet in the beam
splitter used
for the Bell state measurement, their interference is incomplete and
the resulting state is
\begin{eqnarray}
|Q_D\rangle & = &
D(\frac{i\theta}{2\alpha^{2}})|Q\rangle|B_{00}\rangle \nonumber\\
& = &
\mu |-\sqrt{2}\alpha+i\delta,-i\delta,-\alpha\rangle+
\mu |i\delta,\sqrt{2}\alpha-i\delta,\alpha\rangle+ \nonumber\\
& &
\nu |i\delta,-\sqrt{2}\alpha-i\delta,-\alpha\rangle+
\nu |\sqrt{2}\alpha+i\delta,-i\delta,\alpha\rangle
\end{eqnarray}
where $\delta=\frac{\theta}{2\sqrt{2}\alpha}$.  Because the qubit in
mode $a$ was corrupted by the displacement operator, now it is
possible
to detect photons in both modes $a$ and $b$ simultaneously.  We now
detect $n_a$ photons in mode $a$ and $n_b$ photons in mode $b$, and
this measurement leaves mode $c$ in the pure state given by
\begin{multline}
\langle n_a| \langle n_b|Q_D\rangle= \frac{1}{\sqrt{2}} \exp
\left(-\alpha^2-\frac{\theta^2}{8\alpha^2}\right)\frac{1}{\sqrt{n_a!n_b!}}\\
\left(\sqrt{2}\alpha\right)^{n_a+n_b}
[\mu
(-1)^{n_a+n_b}(1-i\epsilon)^{n_a}(i\epsilon)^{n_b}|-\alpha\rangle \\+
\mu (i\epsilon)^{n_a}(1-i\epsilon)^{n_b}|\alpha\rangle\\+
 \nu (-1)^{n_b}(i\epsilon)^{n_a}(1+i\epsilon)^{n_b}|-\alpha\rangle\\+
\nu (-1)^{n_b}(1+i\epsilon)^{n_a}(i\epsilon)^{n_b}|\alpha\rangle],
\end{multline}
where $\epsilon=\frac{\theta}{4\alpha^2}$, and we have ignored the
normalization factor due to the nonorthogonality of the computational
basis states. This state may need to be corrected with $X$ or $Z$
operations and properly normalized before we obtain the final result
of the teleportation, which we will call $|Q_{n_a,n_b}\rangle$.  We
can see that this state is close to our goal by examining the limit
when $\epsilon\ll 1$. In this case we are almost certain to measure
one of $n_a$ or $n_b$ to be zero. The number of photons in the other
mode is given by a probability distribution which is almost exactly
equal to the Poisson distribution with a mean of $2\alpha^2$.  This
leaves us with the state
\begin{eqnarray}
& \approx &
\mu(1-in\epsilon)|-\alpha\rangle+\nu(1+in\epsilon)|\alpha\rangle \\
& \approx & \mu e^{-in\epsilon}|-\alpha\rangle+\nu
e^{in\epsilon}|\alpha\rangle \\
& = &
R\left(Z,\frac{n\theta}{2\alpha^2}\right)\left(\mu|-\alpha\rangle+
\nu|\alpha\rangle\right).
\end{eqnarray}

\begin{figure}
   \begin{tabular}{cc}
\includegraphics[scale=.4]{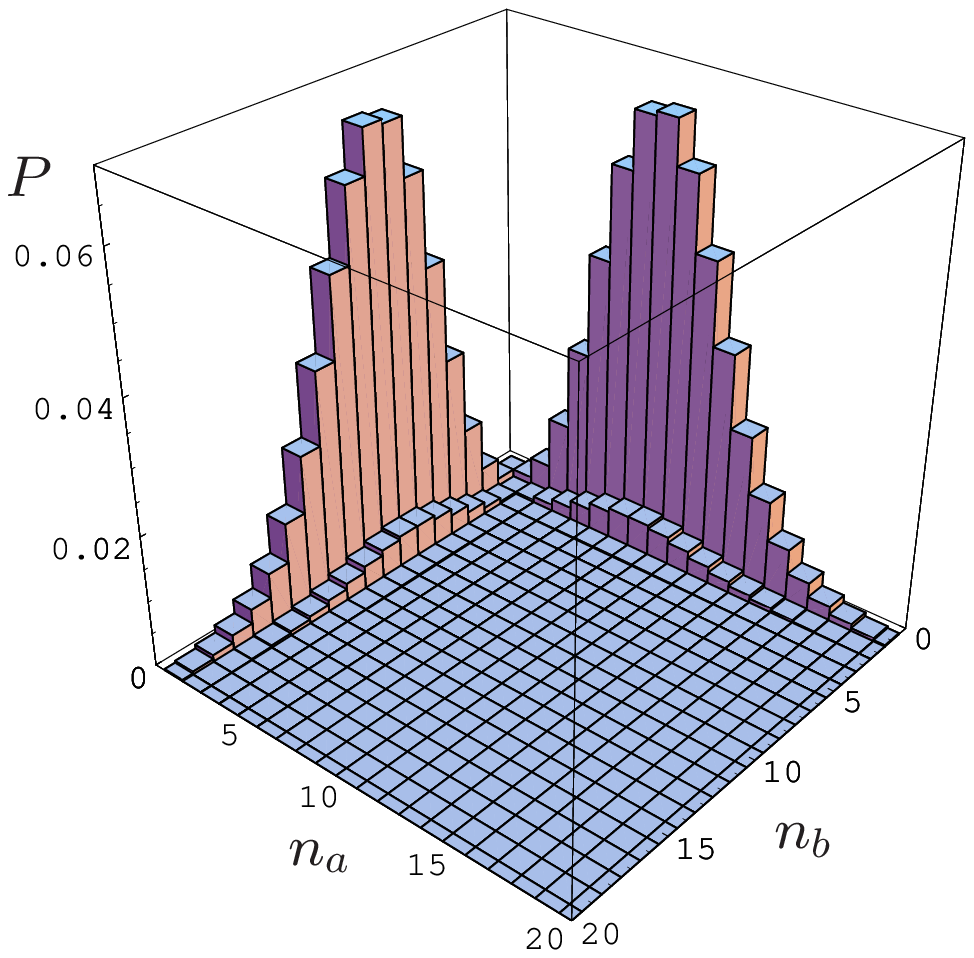} &
\includegraphics[scale=.4]{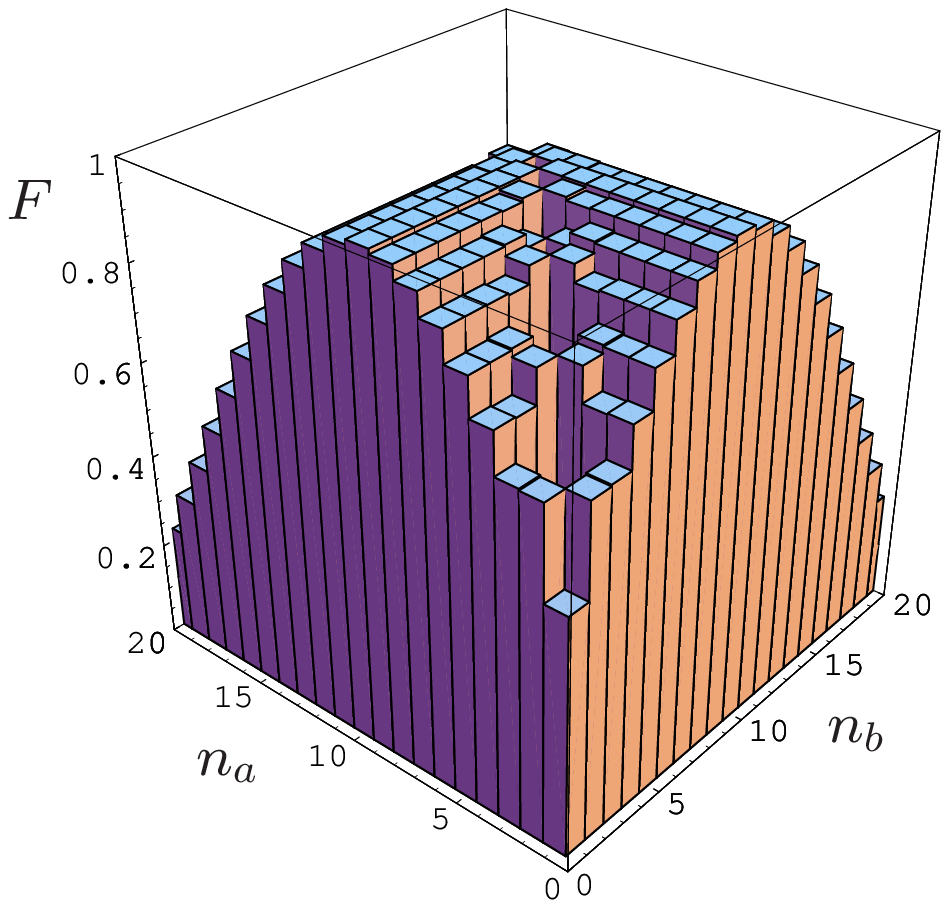} \\
(a) & (b)
\end{tabular}
\caption{Here we plot (a) the probability to detect the pair
   $n_a$ and $n_b$ when performing the $R(Z,\frac{\pi}{2})$ rotation,
   and (b) $F=|\langle Q_{n_a,n_b}|Q_{goal}\rangle|^2$ as a function of
   $n_a$ and $n_b$.  We use the worst case input qubit and an
   $\alpha=2$.}
\label{fig:qandp}
\end{figure}

To evaluate the effectiveness of this procedure without making such
severe approximations, we examine $|Q_{n_a,n_b}\rangle$ in Fig.
\ref{fig:qandp},  where we calculate the fidelity $|\langle
Q_{n_a,n_b}|Q_{goal}\rangle|^2$ and the probability to measure $n_a$
and $n_b$. We use $\alpha=2$, the input qubit
$|Q\rangle=\frac{1}{\sqrt{2}}(|-\alpha\rangle+|\alpha\rangle)$, and a
rotation angle of $\theta=\frac{\pi}{2}$.  These choices for
$|Q\rangle$ and $\theta$ give the worst case scenario, in which we
obtain the lowest fidelity with
$|Q_{goal}\rangle=R(Z,\theta)|Q\rangle$.  Because the $Z$ operation
is equivalent to $R(Z,\pi)$, we can reach any angle by using $Z$ and
$R(Z,\theta)$ where $\theta\leq\frac{\pi}{2}$.  One can see that
$|\langle Q_{n_a,n_b}|Q_{goal}\rangle|^2$ is very close to one in the
regions where we are most likely to detect the pair $n_a$ and $n_b$.

\begin{figure*}
\includegraphics[scale=.9]{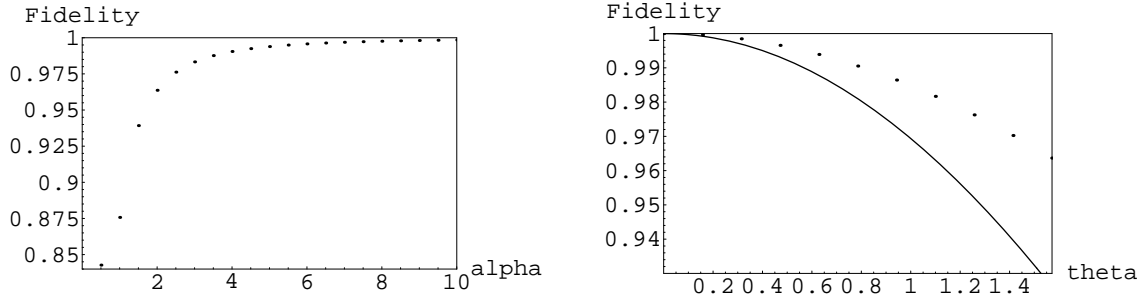}
\caption{Here we plot the fidelity of our procedure for performing
the $R(Z,\theta)$ rotation as a function of $\alpha$ (using
$\theta=\pi/2$) and as a function of $\theta$ (using $\alpha=2$).
The dots show the fidelity after the teleportation and the curve
shows the fidelity before teleportation.  We use the worst case input
qubit.}
\label{fig:alphaandtheta}
\end{figure*}

In order to compute the overall fidelity of this operation, we first
construct the mixed state $\rho$ representing the output of the
teleportation operation for all measurement results
\begin{equation}
\rho=\sum_{n_a=0}^{\infty}
\sum_{n_b=0}^{\infty}P(n_a,n_b)|Q_{n_a,n_b}\rangle\langle
Q_{n_a,n_b}|.
\end{equation}
The fidelity is then given by
\begin{equation}
F=\langle Q_{goal}|\rho|Q_{goal}\rangle.
\end{equation}
We plot $F(\alpha)$ for $\theta=\frac{\pi}{2}$ and $F(\theta)$ for
$\alpha=2$ in Fig. \ref{fig:alphaandtheta}.  We can obtain a fidelity
of 0.99 or above for any desirable angle if we can produce qubits
with $\alpha=5.5$.  A second strategy would be to limit our operation
of $R(Z,\theta)$ to small angles.  Larger rotations could be built
from repeated applications of a high fidelity gate.  For example the
fidelity for $\theta=\pi/16$ is $F=0.99880$ when $\alpha=2$.
Repeating this 8 times implements $R(Z,\pi/2)$ with a fidelity of
$F=0.99880^{8}=0.99044$.  Compare this with the fidelity of $0.92865$
when performing $R(Z,\pi/2)$ in a single step.

\begin{figure}
\includegraphics[scale=.8]{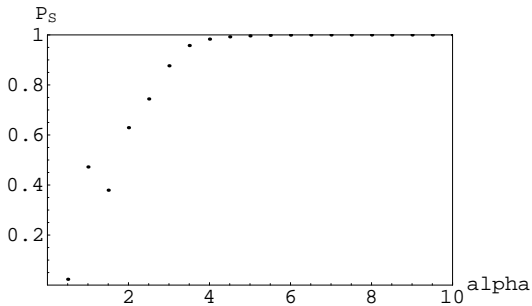}
\caption{Here we plot $P_S(\alpha)$ the probability that our
implementation of $R(Z,\theta)$ succeeds given that we demand it
performs with a fidelity of 0.99. Here again we use $\theta=\pi/2$
and the worst case qubit. Notice that $P_S(1)>P_S(1.5)$, contrary to
what we might expect.  This occurs because when $\alpha=1.5$,
$2\alpha^2$ is not an integer, and the maximum possible $|\langle
Q_{n_a,n_b}|Q_{goal}\rangle|^2$ is therefore significantly lower than
we can find in the $\alpha=1$ case.}
\label{fig:ps}
\end{figure}

Yet a third strategy emerges if we are willing to operate the logic
gate in a non-deterministic fashion, in which the gate sometimes
fails and must be repeated with a new copy of the qubit.  Qubits can
be protected from destruction if we use the gate teleportation scheme
of \cite{gott} as pictured in Fig. \ref{fig:fid210} and discussed in
the previous section.  We can then simply discard $R(Z,\theta)$
attempts for which the measurements of $n_a$ and $n_b$ yield low
values for the product $|\langle Q_{n_a,n_b}|Q_{goal}\rangle|^2$.
Suppose we choose a set $S$ of $(n_a,n_b)$ pairs which are accepted
as successful operations of the logic gate, and $P_S$ is the
probability to measure a member $S$ during the teleportation.   The
total output of the logic gate (when it succeeds) is then the mixed
state
\begin{equation}
\rho_S=\frac{1}{P_S}\sum_{(n_a,n_b)\in S}
P(n_a,n_b)|Q_{n_a,n_b}\rangle\langle Q_{n_a,n_b}|.
\end{equation}
We can now operate this logic gate with a fidelity which is very
close to one.  Of course, this is limited by the maximum possible
value of $|\langle Q_{n_a,n_b}|Q_{goal}\rangle|^2$ (0.999995 for
$\alpha=2$ and $\theta=\pi/2$ with the worst case qubit).  Suppose we
insist on performing $R(Z,\theta)$ with a fidelity of 0.99.  In Fig.
\ref{fig:ps} we plot $P_S$ as a function of $\alpha$ under this
restriction.  This allows us to make estimates of the number of
Bell-cat states required to perform a single $R(Z,\theta)$.  In the
gate teleportation scheme, each attempt to perform $R(Z,\theta)$
requires 2 Bell-cat states, so on average we need $2/P_S$ Bell-cats.
Because there is a 50\% probability of performing $R(Z,-\theta)$,
during the gate teleportation, we need an additional $2/P_S$
Bell-cats to correct this.  Because $Z$ commutes with $R(Z,\theta)$
it is not necessary to perform $Z$ after each teleportation; instead
we can wait and perform only one $Z$ after all teleportations are
complete.  This makes a total of $4/P_S+1$ Bell-cats on average, or
11.55 for $\alpha=1$, or 5.75 for $\alpha=4$.

Which of these three strategies, (i) using very large $\alpha$, (ii)
using only small $\theta$, or (iii) operating the gate
probabilistically and using gate teleportation, is ultimately most
efficient is a complicated question that will depend on the
constraints of Bell-cat production and photon counters.  We hope to
address this further in future research.

The other gates of the previous section can similarly be implemented
by replacing the projective measurements with photon counting
measurements. In this way we are able to implement a universal set of
quantum gates on the coherent state qubits via linear optics, photon
counting and cat and Bell-cat state resources. We now examine how the
cat and Bell-cat states may be produced.

\section{The generation of small Schr\"odinger cats states}

Let us now turn our attention to how small amplitude Schr\"odinger
cat states required for our universal quantum computation schemes can be
realized using technologies currently available or likely in the near future.
More specifically how do we generate states of the form
\begin{eqnarray}\label{eqn:cat}
|\Psi_\pm\rangle&=& \frac{1}{\sqrt{\cal{N}_\pm}} \left[|-\alpha\rangle \pm
|\alpha\rangle\right]
\end{eqnarray}
where the ${\cal N}_\pm=2\pm 2 e^{-2 |\alpha|^2}$.
As we have seen the amplitude of these
cat states need not be large ($\alpha \approx 2$ is sufficient).
An elegant proposal was made by Dakna et.al\cite{dakna97} (see also
\cite{song}) for generating such
states by means of a conditional measurement on a beam splitter. Their scheme
is depicted in Fig (\ref{fig0}) and works as follows:
\begin{figure}[!htb]
\includegraphics[scale=0.25]{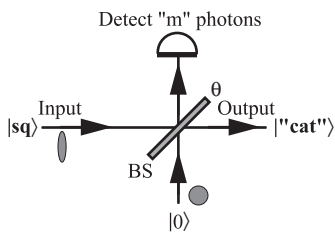}
\caption{Schematic diagram for the generation of a Schr\"odinger like
cat states by means of a conditional photon number measurement
on a beam splitter. A single mode squeezed state is input into one port of
a variable reflectivity beam-splitter with the other input being a
vacuum state.  A definite measurement of $m$ photons (with $m>0$) on
one output port of the beam-splitter prepares to a
good approximation the required state.}
\label{fig0}
\end{figure}
A squeezed state of the form $| \Psi_{sq}\rangle=
\left(1-|\lambda|^2 \right)^{\frac{1}{4}} \sum_n \frac{
\sqrt{(2n)!}}{n!} \left(\frac{\lambda}{2}\right)^n | 2 n \rangle$
(with squeezing parameter $\lambda$) and a vacuum state
$| 0\rangle$ are combined on a variable transmissivity $\theta$ beam-splitter.
On the second output port from the beam-splitter a definite photon number
measurement, which can be modelled
by the POVM $| m\rangle \langle m|$, is performed giving a result $m$.
The conditional state of the remaining output mode is then
\begin{eqnarray}\label{daknacat}
|\Psi_m\rangle&=& \frac{1}{\sqrt{\cal N}_m} \sum_n c_{n,m} \left(
\frac{\lambda \cos^2 \theta}{2}\right)^\frac{n+m}{2} | n
\rangle
\end{eqnarray}
with $c_{n,m}=(n+m)!\left(1+\left(-1\right)^{n+m}\right)/(\sqrt{n!}
\Gamma \left(\frac{n+m}{2}+1\right))$ and
${\cal N}_m= \sum_n c_{n,m}^2 \left| \frac{\lambda \cos^2
\theta}{2}\right|^{n+m}$. The mean photon number for Eq.\ref{daknacat}
is
\begin{eqnarray}
\langle \bar n \rangle=\frac{1}{{\cal N}_m} \sum_n n c_{n,m}^2
\left| \frac{\lambda \cos^2 \theta}{2}\right|^{n+m}
\end{eqnarray}

Eq.\ref{daknacat} can be broken into two cases: the state resulting
from an even $m$ result and the state from an odd $m$ (which will not 
be considered
here). For $m$ even
Eq.\ref{daknacat} has only even photon numbers and can be written
in the simplified form
\begin{eqnarray}\label{evendaknacat}
|\Psi_m\rangle&=& \frac{1}{\sqrt{\cal N}_m} \sum_n \frac{(2 n+m)!
\left(\frac{\lambda \cos^2 \theta}{2}\right)^{n+\frac{m}{2}} }
{ \left(n+\frac{m}{2}\right)! \sqrt{(2n)!}} | 2 n \rangle
\end{eqnarray}
For $\lambda \cos^2 \theta$ small, this expression can be furhter 
approximated as
\begin{eqnarray}
|\Psi_m\rangle&\approx& | 0 \rangle  + \lambda \cos^2 \theta
\frac{1+m}{\sqrt{2}}| 2 \rangle+\ldots
\end{eqnarray}
Here we observe that as $m$ increases so does the population in the 
$| 2 \rangle$
(and higher) states compared with the $m=0$ situation. Thus for small
$\lambda \cos^2 \theta$ the mean photon number increases as $m$ increases.
As a cautionary note, we must emphasize that the scheme here requires the
detection of an exact number of photons to generate the approximate single
mode cat state. Currently detectors are not that efficient but good
progress is being made.

Here the resulting

\begin{figure}[!htb]
\begin{center}
\includegraphics[scale=0.5]{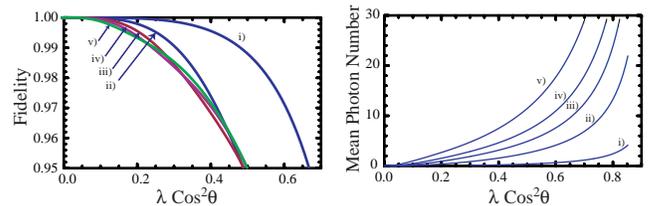}
\end{center}
\caption{
Plot of the fidelity of the state Eq.\ref{daknacat} compared with
Eq.\ref{eqn:cat} and mean photon number of Eq.\ref{daknacat}
versus $\lambda \cos^2 \theta$
for i) m=0, ii) m=2, iii) m=4, iv) m=6, and v) m=10.}
\label{fig2}
\end{figure}

\begin{figure}[!htb]
\includegraphics[scale=0.35]{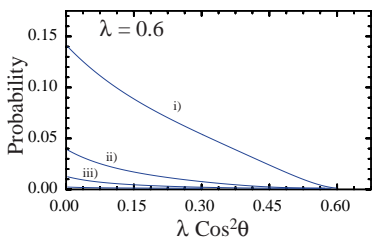}
\caption{Plot of the probability of generating Eq.\ref{daknacat}
versus $\lambda \cos^2 \theta$ for $\lambda=0.6$ with
i) m=2, ii) m=4, iii) m=6, and iv) m=10.}
\label{fig3}
\end{figure}

Now let us determine how good an approximation Eq.\ref{daknacat} is with
the Schr\"odinger cat states given by Eq.\ref{eqn:cat}. This can be achieved
by calculating the overlap $F=|\langle\Psi_+ |\Psi_m\rangle|^2$ between the
two states. To this end we plot in Fig (\ref{fig2})
both the mean photon number of the state
of Eq.\ref{daknacat} and the fidelity for various even $m$.
It is interesting to observe that a good fidelity ($>$95\%) can be achieved
for quite a range of $\lambda \cos^2 \theta$ and $m$. In fact for
$\lambda \cos^2 \theta \leq 0.3$ the fidelity between the two states we are
comparing exceeds 99\%. However to achieve a cat state with a moderate
mean photon number we either need $m$ large or $\lambda \cos^2 \theta 
\geq 0.5$.
As $m$ increases the overlap between Eq.\ref{eqn:cat} and Eq.\ref{daknacat} for
the same mean photon number increases. There is a potential regime
where Eq.\ref{daknacat} has moderate mean photon number and a high overlap with
the state Eq.\ref{eqn:cat}. However there is a trade off in that the 
initial probability
of generating the state Eq.\ref{daknacat} with $\lambda$ fixed decreases as $m$
increases. The probability of successfully generating the state
Eq.\ref{daknacat} is given by
\begin{multline}\label{probdaknacat}
{\rm P}_m = \sqrt{\frac{1-\lambda^2}{1-\lambda^2 \cos^4
\theta}}\left[\frac{\lambda^2 \sin^2 2\theta} {4 \left(1-\lambda^2
\cos^4 \theta\right)} \right]^m \\ \sum_{l=0}^{\rm Int[m/2]} 
\frac{m!}{(m-2 l)!l!^2
(2 \lambda \cos^2 \theta)^{2l}}
\end{multline}
and is shown in Fig (\ref{fig3}) for various $m$.
As $m$ increases the probability of successfully generating our required
state significantly decreases but the success probability is reasonable for
$\lambda=0.6$ with either $m=2$ or $4$. With such parameters we can generate a
Schr\"odinger cat like state with a fidelity greater than 95\%
with a probability of success greater than one percent.

\begin{figure}
\includegraphics[scale=0.5]{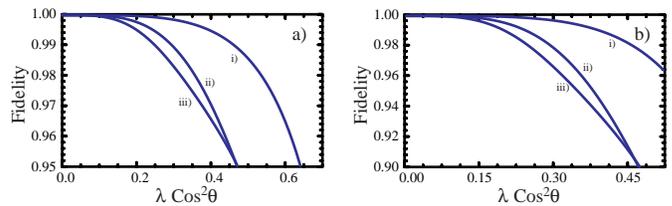}
\caption{Plot of the fidelity for the a) dakna two mode cat state
versus $|\alpha\rangle|\alpha\rangle+ |-\alpha \rangle |-\alpha\rangle$
and b) the state $e^{i \phi} |\alpha\rangle +e^{-i \phi} |-\alpha\rangle$
resulting from the action of the gate $R(Z,\phi)$ with $\phi=\pi / 32$
versus $\lambda \cos^2 \theta$ for i) m=0, ii) m=2 and iii) m=4.}
\label{fig5}
\end{figure}

Let us now determine if the Dakna cat state can be used to generate 
the entangled
cat state $|\alpha\rangle|\alpha\rangle+ |-\alpha \rangle |-\alpha\rangle$
required in the teleportation step of the various fundamental gates. 
Such a state can
be generated from an ideal cat state but combining it with the vacuum 
state on a 50/50
beamsplitter (here we need to choose the amplitude $\beta$ of the 
original single mode
cat to be $\sqrt{2} \alpha$. Using the Dakna state cat as the input to this
beamsplitter, we plot in  Figure (\ref{fig5}a) the overlap
between the resulting two mode state and the two mode entangled
state
We observe that for both $m=2,4$ we have the fidelity exceeding
ninety five percent for a wide range of parameters. This indicates
that to a very good approximation we can generate the two mode
entangled cat state required for our basic gate operations.
Given this entangled resource we can now investigate one such gate
operation. We consider the operation of the $R(Z,\phi)$ gate
illustrated in Fig (\ref{fig:fid210}) using the Dakna cat state to generate
both the entangled resource and the state $|Q\rangle$. In
Fig (\ref{fig5}b) we show the fidelity for performing the gate operation to
transform the state $|Q\rangle$ to $e^{i \phi} |\alpha \rangle+ ^{-i \phi}
|-\alpha \rangle$ for $\phi$ small. This results show the feasibility of
performing in principle experiments to demonstrate quantum logic.

\section{Error Correction}

A viable quantum computation scheme must be capable of incorporating
error correction. We now briefly discuss the issue of error
correction.
The major sources of error in our scheme are expected to
be, in order of
increasing significance:
(i) errors due to non orthogonal code states;
(ii) errors due to moving outside the qubit basis;
(iii) errors due to random optical phase shifts and;
(iv) photon loss.

Sources (i) and (ii) are equivalent. As discussed in section II we
could use as orthogonal code states the cat states. These states are
a single qubit gate away from the coherent state code. Such a gate
must be non unitary and we have given a method based on teleportation
to achieve this. Single qubit manipulations in the cat state basis
require us to move outside of the qubit basis and rely on
teleportation to project back into the computational basis. We have
shown that errors introduced in this process due to non orthogonality
of coherent states are exponentially small in the amplitude and in
any case are heralded by the teleportation process itself. If we see
an error we can repeat the teleportation process which, as the errors
can be made so small, is very likely to succeed after a couple of
trials. We will thus not consider these sources of error further.

Optical phase shift errors will occur due to timing errors between
different qubits and between qubits and the local oscillator. Such
errors may arise from path length fluctuations in the circuit. These
can be monitored and corrected through classical optical
interferometric techniques. Such locking techniques are a mature
technology and can be extremely precise. We will assume sufficient
classical control is exercised to make these errors negligible.

Photon loss error however is a more serious problem as it is
never heralded and increases
quadratically with $\alpha$. In this case we must turn to error
correction coding to mitigate the effect.  Photons are lost from a
coherent state at Poisson distributed times at a rate determined by
$\gamma\langle a^\dagger a\rangle$, where $\gamma$ is the single
photon loss rate. Obviously, if a photon is lost the system has one
less photon. The effect of photon loss from a pure state is thus
given by  $|\psi\rangle \to a |\psi\rangle$, where $a$ is the Bose
annihilation operator.

The Poisson distributed nature of photon
loss means that even when no photons are lost from a coherent state,
the state must change.  Not seeing a photon emitted up to time $t$
indicates that the state is increasingly likely not to contain any
photons at all and thus we must continuously adjust our description
of the state to reflect this knowledge.

We can put the description
of photon loss on a more formal basis by asking for the conditional
state of the system given an entire history of photon loss events.
This is a list of times $\{t_1< t_2, < \ldots< t_n\leq  t\}$, at
which photons are lost. The (unnormalized) conditional state
is\cite{SrinivasDavies}

\begin{multline}
|\psi(t|t_1,t_2,\ldots,t_n)\rangle=\gamma^{n/2}e^{-\gamma(t-t_n)/2}a
e^{-\gamma(t_n-t_{n-1})/2}a\\ \ldots a e^{-\gamma(t_2-t_1)/2}a
e^{-\gamma t_1/2}|\psi(0)\rangle
\end{multline}
The norm of this
unconditional state is the probability for this history.

If we
start in the  coherent state $|\alpha\rangle$  and lose no photons up
to time $t$, the conditional state is $|\kappa \alpha\rangle$ where
$\kappa= e^{-\gamma t/2}$. The important fact here is that the state
remains a coherent state even though the amplitude is decreased. This
kind of error takes us out of the code space, but can be corrected by
teleportation. Consider the
state
\begin{equation}
|\Psi\rangle=(\mu|-\kappa\alpha\rangle_1+\nu|\kappa\alpha\rangle_1)
(|\alpha,\alpha\rangle_{23}+|-\alpha,-\alpha\rangle_{23})
\end{equation}
If
we mix modes $1,2$ on a beam splitter, and count $n\neq 0$ photons in
mode 1 and $0$ photons in mode 2, the conditional state of mode 3 is
found to be $\mu|-\alpha\rangle+\nu|\alpha\rangle$.  If $\kappa$ is
small enough this will occur with high probability. In fact letting
$\kappa=1-\epsilon$ the probability for this event is very close
to
\begin{equation}
P(n_1\neq
0,n_2=0)=e^{-\epsilon^2|\alpha|^2/2}
\end{equation}
the teleportation
projects us back into the qubit basis with high probability as it is
most likely that $n_1$ is near $2|\alpha|^2$.  Failure of the
protocol is heralded by $n_1=0,n_2\neq 0$ and thus the gate can be
repeated if necessary. The dominant term in the failure probability
is approximately given $e^{-2|\alpha|^2}$. In fact this resetting of
the amplitude happens as a matter of course in all the teleportation
based gates we have discussed. Thus it may not be necessary to
explicitly introduce additional gates for this purpose.

If  a photon is
lost  from  a coherent state, the state is unchanged up to a phase as
$a|\alpha\rangle=\alpha|\alpha\rangle$, which when normalized
produces only a phase shift given by the phase of $\alpha$ \cite{coc98}. This
means that, in the qubit code space, photon loss is equivalent to an
erroneous  application of the $Z$ gate, which induces a sign flip error.
A sign flip error may be converted into a bit flip error by
performing a Hadamard gate and working in the conjugate basis
$|\pm\rangle=|\alpha\rangle\pm|-\alpha\rangle$, ( that is the cat
states). To prepare a code state to protect sign flip errors, we thus
first prepare  the standard three qubit
code\cite{got96},
\begin{equation}
|0\rangle_L=|-\alpha,-\alpha,-\alpha\rangle\
\ \ ;\ \ \
|1\rangle_L=|\alpha,\alpha,\alpha\rangle
\end{equation}
and then
perform a Hadamard gate on each mode separately. Sign flip errors
will now appear as bit flip errors and can be corrected using the
standard three qubit circuit \cite{NielChu}.

The encoding is easily
done in linear optics by an extension of the technique previously
discussed for producing Bell entanglement.
Two beam splitters suffice to implement the
transformation
\begin{multline}
(\mu|-\beta\rangle_1+\nu|\beta\rangle_1)|0\rangle_2|0\rangle_3\to\\
\mu|\frac{-\beta}{\sqrt{3}}\rangle_1
|\frac{-\beta}{\sqrt{3}}\rangle_2 |\frac{-\beta}{\sqrt{3}}\rangle_3
+\nu|\frac{\beta}{\sqrt{3}}\rangle_1
|\frac{\beta}{\sqrt{3}}\rangle_2 |\frac{\beta}{\sqrt{3}}\rangle_3
)
\label{three-cat}
\end{multline}
At the first beam splitter, with
reflectivity amplitude of $\frac{1}{\sqrt{3}}$, modes 1 and 2 are
combined, subsequently modes 2 and 3 are combined at a $50/50$ beam
splitter.
Thus by choosing $\beta=\sqrt{3}\alpha$ we can
immediately prepare the entangled state
$\mu|-\alpha,-\alpha,-\alpha\rangle+\nu|\alpha,\alpha,\alpha\rangle$.

Any logical operation may be performed on an
arbitrary state in the code
space
\begin{equation}
|\psi\rangle_L=\mu|-\alpha,-\alpha,-\alpha\rangle+
\nu|\alpha,\alpha,\alpha\rangle
\end{equation}
by
extending the teleportation gates for the single mode case to the
multi mode case.  Displacements can easily be done one mode at a
time. The teleportation steps in the gates
will require a six mode entangled resource of the
form
\begin{equation}
|\alpha,\alpha,\alpha,\alpha,\alpha,\alpha
\rangle
+|-\alpha,-\alpha,-\alpha,-\alpha,-\alpha,-\alpha \rangle
\end{equation}
Such a state could be prepared  by an obvious
generalization of the method used in Eq.(\ref{three-cat}), however
the amplitude of the initial cat state is becoming
uncomfortably large. We now show how to avoid this problem.

Consider the resource state
\begin{equation}
      |-\alpha, -\sqrt{2} \alpha \rangle + |\alpha, \sqrt{2} \alpha \rangle
\end{equation}
which can be produced from a cat state of amplitude $\sqrt{3} \alpha$
by splitting it on a beamsplitter of reflectivity $1/\sqrt{3}$.
Suppose this state is used as the entanglement in a teleportation
protocol with the smaller amplitude arm being mixed with the input
state and measured. The result of the teleportation is the
transformation
\begin{equation}
\mu|-\alpha\rangle+
\nu|\alpha\rangle \to \mu|-\sqrt{2} \alpha\rangle+
\nu|\sqrt{2} \alpha\rangle
\end{equation}
where we have assumed the necessary bit-flip and sign-flip
corrections have been made.
That is, the state is amplified whilst preserving the superposition.
If the amplified state is then split on a 50:50 beamsplitter an
entangled state of the same amplitude as the original will be produced.
By repeating this process many times multi mode encoded states or
entangled resource states can be produced deterministically without the need to
produce ``large'' cats.

Finally we note that the preceding analysis has ignored the effect
of gate errors due to photon loss. For the phase rotation gate and the
control phase gate ($R(Z,\theta)$ and $R(Z\otimes Z,-\phi)$) the
effect of photon loss is similar to that discussed above for the
propagating qubit, that is it produces sign flips in the computational
basis. In reaching this conclusion we have considered loss events
occurring; to the resource states; at the measurement site and; at the
displacement. Hence errors in these gates can be corrected by the code
discussed above. However photon
loss events in the superposition gate ($R(X,\pi/2)$) can produce
bit-flips in the computational basis if they occur at the measurement
site. As a result, protecting a general circuit will require error
correction for both sign flips and bit flips. This can be achieved by
using the standard nine-qubit code \cite{NielChu} which can be
implemented by a straightforward generalization of the techniques
outlined in the preceding discussion.

It is likely that the application of more efficient codes \cite{laf}
and optimization, in particular exploiting the rarity of
bit flip versus sign flip errors in a general circuit, can reduce the
complexity of the required error correcting codes. We leave an
investigation of this and the general question of fault tolerance
levels for future research.

\section{Conclusion}

In this paper we have presented a quantum computation scheme based
on encoding qubits as coherent states of equal absolute amplitude but
opposite sign.
The optical networks required to manipulate the qubits
are conceptually simple and require only
linear interactions and photon counting, provided coherent
superposition ancilla states are available (cat states). We have shown
that qubits with amplitude $|\alpha|=2$ and resource cat states of
amplitude $|\alpha|=\sqrt{6}$ would be sufficient. Accurate photon
counting measurements of up to about 10 photons would also be necessary.

We have discussed how the cat-state resources could
be produced from squeezed sources, linear
interactions and photon counting in a simple scheme. This scheme
appears capable of producing states suitable for proof of principle
experiments. It seems likely though that more sophisticated schemes
would be necessary for scalable systems.

The power of the scheme stems from the ability to generate
entangled states and make Bell basis measurements with simple
linear interactions. This means teleportation protocols of various
forms can be implemented deterministically to great effect.

A disadvantage of the scheme is that the multi-photon nature
of the qubits make them more susceptible to photon loss than single
photon qubits. However, we
have shown how error correction can be employed in a straightforward
way to counter this effect.

Being a simple optical system, the
decoherence and control issues are well understood and with sufficient
effort realistic
evaluations of the resources and precision needed can be made. This
level of understanding is not a feature of all quantum computer
candidates. As well as the long term
goal of quantum computation, nearer term applications in quantum communication
protocols appear possible.

SG acknowledges support from the Arthur J. Schmitt Foundation and thanks
John LoSecco and Hilma Vasconcelos for valuable discussions. WJM
acknowledges funding in part by the European project
EQUIP(IST-1999-11053). AG was supported by the New Zealand Foundation
for Research, Science and Technology under grant UQSL0001.The
Australian Research Council and ARDA
supported this work.

%


\begin{thebibliography}{99}

%
%

\bibitem{KLM} E.~Knill and L.~Laflamme and G.~J.~Milburn,
Nature {\bf 409}, 46 (2001).

\bibitem{RMM} T.~C.~Ralph, W.~J.~Munro and G.~J.~Milburn,
Proceedings of SPIE {\bf 4917}, 1 (2002); quant-ph/0110115.

\bibitem{Bra98}S.~L.~Braunstein and H.~J.~Kimble, \prl {\bf 80}, 869
(1998).

\bibitem{llo} S.~Lloyd and S.~L.~Braunstein, Phys Rev Lett {\bf 82},
1784 (1999).

\bibitem{sand} S.~D.~Bartlett,
Hubert de Guise and B.~C.~Sanders, \pra {\bf 65}, 052316 (2002).

\bibitem{kim01} H.~Jeong and M.~S.~Kim, \pra {\bf 65}, 042305 (2002).

\bibitem{pre} D.~Gottesman, A.~Kitaev and J.~Preskill, Phys Rev A {\bf
64},
012310 (2001).

\bibitem{WallsMilb94} D.~F.~Walls and G.~J.~Milburn, {\it Quantum
Optics}
(Springer-Verlag, Berlin, 1994).

\bibitem{note1} ``Taking $\alpha$ real'' means that the field is in
phase with the local oscillator which is used for qubit measurement
and to make the
displacements
required for some of the
gates. The average intensity of all logical pulses is $I=\hbar
\omega |\alpha|^{2}$ per bandwidth with $\omega$ the optical
frequency.

\bibitem{bennett} C.~H.~Bennett, G.~Brassard, C.~Crepeau, R.~Jozsa,
A.~Peres and W.~K.~Wootters, \prl, {\bf70}, 1895 (1993).

\bibitem{zeno} B.~Misra and E.~C.~G.~Sudarshan, J. Math. Phys. {\bf 18},
756 (1977).

\bibitem{gott} D.~Gottesman and I.~L.~Chuang, Nature {\bf 402}, 390
(1999). M.~A.~Nielsen and I.~L.~Chuang, Phys Rev Lett {\bf 79}, 321
(1997).

\bibitem{reid} L.~Krippner, W.~J.~Munro, and M.~D.~Reid, Phys Rev A
{\bf 50},
4330 (1994).

\bibitem{enk02} S.~J.~van Enk and O.~Hirota, \pra {\bf 64},
022313 (2001).

\bibitem{kim2} H.~Jeong, M.~S.~Kim, and J.~Lee
\pra {\bf 64}, 052308 (2001).

\bibitem{dakna97}  M.~Dakna {\it et al},
T.~Anhut, T.~Opatrny, L.~Knll and
D.~-G.~Welsch,
Phys Rev A {\bf 55}, 3184 (1997).

\bibitem{song} S.~Song, C.~M.~Caves and B.~Yurke,
Phys Rev A {\bf 41}, 5261 (1990).

\bibitem{SrinivasDavies}M.D.Srinivas adn E.B.Davies, opt.
Acta {\bf 28}, 981 (1981).

\bibitem{coc98} P.~Cochrane, G.~J.~Milburn and W.~J.~Munro, Phys Rev
A {\bf
59}, 2631 (1998).

\bibitem{got96} D.~Gottesman, Phys Rev A {\bf 54}, 1862 (1996),
D.~Gottesman, Phys Rev A {\bf 57}, 127 (1998).

\bibitem{NielChu} M.~Nielsen and I.~Chuang,
{\it Quantum computation and quantum information}
(Cambridge University Press, Cambridge, UK 2000).

\bibitem{laf} R.~Laflamme, C.~Miquel, J.~P.~Paz, and W.~H.~Zurek,
\prl {\bf 77}, 198 (1996).
%
%
%
%
%
%
%

\end{thebibliography}
\end{document}